\documentclass[
aip,jcp,amsmath,amssymb,reprint,
]{revtex4-1}

\usepackage[english]{babel}
\usepackage{placeins}
\usepackage{float}
\usepackage[normalem]{ulem}
\usepackage{graphicx}
\usepackage{xargs} 
\usepackage[pdftex,dvipsnames]{xcolor} 
\usepackage{siunitx}
\usepackage{tikz}
\usepackage{svg}
\usepackage[colorlinks=true, allcolors=blue]{hyperref}
\usepackage[capitalize]{cleveref}
\usepackage{xr}

\newcommand{\DLR}{\affiliation{Institute of Engineering Thermodynamics, German Aerospace Center (DLR), Wilhelm-Runge-Str. 10, 89081 Ulm, Germany}}
\newcommand{\HIU}{\affiliation{Helmholtz Institute Ulm, Helmholtzstr. 11, 89081 Ulm, Germany}}

\makeatletter
\def\@email#1#2{%
 \endgroup
 \patchcmd{\titleblock@produce}
  {\frontmatter@RRAPformat}
  {\frontmatter@RRAPformat{\produce@RRAP{*#1\href{mailto:#2}{#2}}}\frontmatter@RRAPformat}
  {}{}
}%
\makeatother
\begin{document}
\author{Konstantin Lamp}
\DLR \HIU
\author{Alejandro D. Somoza}
\DLR \HIU
\author{Elias Walter}
\DLR \HIU
\author{Marina Walt}
\author{Michael Marthaler}
\affiliation{HQS Quantum Simulations GmbH, Rintheimer Straße 23, 76131 Karlsruhe, Germany}
\author{Maria Fernanda Juarez}
\email{fernjuar@gmail.com}
\DLR \HIU
\author{Birger Horstmann}
\email{birger.horstmann@dlr.de}
\DLR \HIU
\affiliation{University of Ulm, Albert-Einstein-Allee 47, 89081 Ulm, Germany}

\title{Optimizing Subspace Expansion in Quantum Chemistry through Operator Selection and Reference State Choice}

\begin{abstract}
The Virtual Quantum Subspace Expansion (VQSE) extends the Variational Quantum Eigensolver (VQE) by leveraging additional measurements on the reference state to capture the influence of excluded virtual orbitals. 
This makes VQSE attractive for chemical applications where accurate energy differences along potential energy surfaces are crucial for modeling reaction rates and kinetics. 
In this work, we analyze VQSE performance on H$_2$ dissociation including references that use Hartree--Fock molecular orbitals with broken spin symmetry. We identify two mechanisms which affect accuracy: overlap of the reference state with the exact full configuration interaction (FCI) wavefunction and operator pool expressivity.
We show these mechanisms are strongly co-dependent. When operators are restricted to act only from the active to the virtual space, results become highly sensitive to the reference, and enlarging the active space does not guarantee improved accuracy. In this case, prioritizing reference overlap over energy minimization is therefore essential.
Adding single excitations and number operators within the active space recovers the accuracy of MR-CISD (multi-reference configuration interaction singles and doubles) regardless of the reference.
In our noisy hardware experiments, we achieve chemical accuracy by adding additional operators and using strict regularization. These findings motivate careful co-design of reference fidelity, pool expressivity, and hardware constraints for practical VQSE deployment.
\end{abstract}

\maketitle

\section{Introduction} 

Quantum computing offers a paradigm for addressing electronic structure problems intractable to classical computers \cite{cao_quantum_2019}. Quantum devices naturally encode many-body wave functions, facilitating the simulation of strongly correlated systems \cite{cao_quantum_2019}. Accurate potential energy surfaces are essential for predicting reaction dynamics and chemical kinetics, since reaction rates depend sensitively on precise energy differences along these surfaces, particularly during bond dissociation where static correlation dominates \cite{boguslawski_orbital_2013}.

The Variational Quantum Eigensolver (VQE) has emerged as a cornerstone hybrid quantum-classical algorithm for estimating molecular ground-state energies on noisy intermediate-scale quantum (NISQ) devices. However, its practical applicability in quantum chemistry is often constrained by circuit depth limitations, barren plateaus in the optimization landscape, and sensitivity to hardware noise \cite{tilly_variational_2022}. To address these challenges, subspace-expansion methods have been developed as post-processing frameworks that leverage additional measurements on a reference wavefunction (typically obtained from a converged VQE calculation), to systematically improve accuracy without increasing quantum circuit complexity.

The Quantum Subspace Expansion (QSE) method, introduced by McClean et al. \cite{mcclean_hybrid_2017}, projects the variational state onto a small subspace spanned by excited determinants or excitation operators applied to the reference. This approach not only enables access to excited-state properties but was also argued to possess intrinsic error-mitigation qualities. Because the generalized eigenvalue problem is solved within a subspace spanned by physically motivated excitation operators, QSE can suppress contributions from noise-induced components that lie outside the target eigenspace, effectively filtering errors via subspace projection and classical diagonalization \cite{mcclean_hybrid_2017,yoshioka_generalized_2022}.
The Virtual Quantum Subspace Expansion (VQSE) extends this paradigm by incorporating excitations into virtual orbitals, thereby expanding the effective active space without requiring additional physical qubits \cite{takeshita_increasing_2020,urbanek_chemistry_2020}. This enables chemically accurate simulations of systems that would otherwise demand significantly larger quantum registers. Other forms of QSEs have been developed, such as quantum equation of motion \cite{ollitrault_quantum_2020}, Krylov space methods \cite{kirby_exact_2023, zhang_measurement-efficient_2024} and methods based on quantum time evolution \cite{parrish_quantum_2019, stair_multireference_2020, klymko_real-time_2022}.

Despite these advantages, practical implementations of QSE and VQSE face a fundamental bottleneck: the generalized eigenvalue problem (GEVP), whose matrices are constructed from quantum measurements, becomes increasingly ill-conditioned as the subspace dimension grows. Statistical noise from finite sampling, combined with near-linear dependencies among basis states, can lead to numerical instabilities that corrupt energy estimates \cite{kwao_generalized_2026}. Consequently, the strategic selection of the excitation pool (the set of operators used to generate the subspace) has emerged as the critical design choice balancing expressivity against stability. Recent work has addressed this challenge through adaptive excitation selection \cite{huang_quantum_2023}, regularization techniques for the overlap matrix \cite{nakamura_adaptive_2024}, and partitioned subspace constructions that iteratively build well-conditioned bases \cite{oleary_partitioned_2025}.

Moreover, the measurement overhead associated with constructing the GEVP matrices scales unfavorably with subspace size and system dimension. Innovative approaches leveraging classical shadows \cite{fischer_large-scale_2025}, locally-biased measurement strategies \cite{nakamura_adaptive_2024}, and resource-efficient generalized QSE formulations \cite{yang_resource-efficient_2025} have recently demonstrated substantial reductions in sampling requirements while maintaining accuracy. These advances, combined with noise-agnostic error mitigation protocols that unify QSE with purification-based techniques \cite{ohkura_leveraging_2023}, suggest that subspace methods can remain viable even in the presence of hardware noise.

Following these QSE developments, we use H$_2$ dissociation as a minimal benchmark to test the VQSE method on current hardware. In line with standard quantum chemistry practice, we construct an orthonormal molecular orbital basis using the Hartree--Fock (HF) method. This method provides a mean-field approximation to the many-body Schrödinger equation, representing the wavefunction as a single Slater determinant of molecular orbitals that minimize the ground-state energy \cite{helgaker_molecular_2014,szabo_modern_1996}. While the original HF formulation imposed only orthonormality \cite{fock_naherungsmethode_1930}, requiring additional constraints from approximate solutions leads to what Löwdin termed the ``symmetry dilemma'' \cite{lefebvre_aspects_1969}: enforcing symmetries raises the variational energy, while relaxing them lowers energy at the cost of losing good quantum numbers. This dilemma is systematically organized in Fukutome's classification \cite{fukutome_unrestricted_1981}: restricted (RHF/ROHF) methods enforce spin-symmetry conservation for both $\hat{S}^2$ and $\hat{S}_z$; unrestricted HF (UHF) conserves $\hat{S}_z$ while breaking $\hat{S}^2$ invariance; and generalized HF (GHF) lifts all spin constraints, permitting orbital spin mixing. 
Strict symmetry preservation maintains good quantum numbers but fails size-consistency at dissociation, i.e.\ that the energy of a dissociated molecule is not equal to the summed energy of the dissociation fragments, while symmetry relaxation enables correct dissociation limits at the cost of spin contamination \cite{jimenez-hoyos_generalized_2011}. The singlet--triplet instability \cite{cizek_stability_1967} exemplifies this trade-off, as can be seen for H$_2$ with RHF energy rising above the triplet state beyond $\sim1.7$~\AA.
Despite these well-established considerations, orbital and reference state choices remain largely unexplored in quantum subspace methods \cite{yoshioka_variational_2022, huang_simulating_2022, motta_quantum_2023, castellanos_quantum_2023, huang_quantum_2023, gandon_nonadiabatic_2024}.

To address this, we compare reference states obtained from different post-Hartree--Fock methods constructed upon orbitals from distinct Hartree--Fock variants, which differ in their spin symmetry conservation. This allows us to benchmark how the choice of molecular orbitals and reference calculation influences the accuracy and stability of the VQSE method across different operator pools for our minimal example.

The rest of the manuscript is structured as follows: In Sec.~\ref{sec:theory} we cover the different Hartree--Fock formalisms and details about the VQSE construction like operator pools and reference states.
In Sec.~\ref{sec:results:classical}, we analyze classical benchmarks to highlight that while broken-symmetry reference states provide improved \textit{a priori} energies, they exhibit reduced overlap with the exact FCI wavefunction, an important implication of the symmetry dilemma. We demonstrate in Sec.~\ref{sec:results:noiseless} that restricting operators to virtual orbitals renders energy estimates highly sensitive to this overlap, whereas including single excitations and number operators within the active space recovers MR-CISD accuracy regardless of the reference state. Furthermore, we show that the total spin expectation value $\langle S^2 \rangle$ serves as an effective diagnostic for solution quality when spin contamination is permitted. In Sec.~\ref{sec:results_vqse_real}, we extend this analysis to noisy hardware, demonstrating that further including additional operators effectively suppresses noise-induced intruder states. We conclude in Sec.~\ref{sec:discussion_conclusion} and motivate future work.
 
\section{Theory} 
\label{sec:theory}
\subsection{Conventional Methods}
\subsubsection{Hartree--Fock Theory}

Hartree--Fock (HF) is an ab-initio methodology to find a mean-field solution of the time-independent Schrödinger equation with an antisymmetric wavefunction consisting only of a single Slater determinant \cite{helgaker_molecular_2014,szabo_modern_1996}.
It is therefore an approximation to an eigenstate of the Hamiltonian. We are considering the electronic degrees of freedom in atomic units using the Born-Oppenheimer approximation. The Hamiltonian reads
\begin{equation}
\hat H = -\frac{1}{2}\sum_{i} \nabla_i^2 - \sum_{i,A}
\frac{Z_A}{r_{iA}} + \sum_{i>j} \frac{1}{r_{ij}},
\label{ham1st}
\end{equation}
where $i$ and $A$ denote electrons and nuclei, $r_{ij} = \vert \vec {r_{i}} - \vec {r_{j}} \vert$ ($r_{iA} = \vert \vec {r_{i}} - \vec {r_{A}} \vert$) distances between electrons (electrons and nuclei), respectively, and $Z_A$ nuclear charges \cite{helgaker_molecular_2014,szabo_modern_1996}.

For practical purposes, the Hamiltonian is typically expanded in a fixed number of basis functions, which implies
that any expectation value evaluated on the approximated wavefunction is limited to the chosen basis set. The true expectation value is only obtained when the basis spans the part of the Hilbert space containing the target wavefunction which is only guaranteed in the limit of an infinitely large basis \cite{helgaker_molecular_2014,szabo_modern_1996}.

In molecular problems, the basis functions of choice are usually hydrogen-like atomic orbitals, where the radial part is often approximated as a linear combination of Gaussian functions \cite{helgaker_molecular_2014,szabo_modern_1996}.
Within the HF method, 
one determines a basis transformation of these initial atomic orbitals that minimizes
the molecular ground state energy, with the constraint that they must form an orthonormal basis \cite{helgaker_molecular_2014,szabo_modern_1996}. This results in a set of optimized molecular orbitals (MO) $\psi_i$, which are linear combinations of atomic orbitals (AO) $\phi_{\mu}$,
\begin{equation}
\psi_i(\vec x_1) = C_{\mu i} \phi_{\mu} (\vec x_1),
\label{MO}
\end{equation}
with coefficients $C_{\mu i}$,
where $\vec {x_i}$ is a spatial and spin coordinate with $\vec {x_i} = (\vec {r_i}, s_i)$ \cite{helgaker_molecular_2014,szabo_modern_1996}. Here and throughout the text we use the Einstein summation convention.

The ground-state HF wavefunction for a system with $N$ electrons is a Slater determinant of $N$ occupied molecular orbitals $\psi_i$. 
The Hamiltonian \eqref{ham1st} in second quantization expressed in the MOs reads
\begin{equation}
\hat H = h_{pq} \hat a_p^\dagger \hat a_q +\frac{1}{2} h_{pqrs} \hat a_p^\dagger \hat a_q^\dagger \hat a_r \hat a_s \,,
\label{eq:ham2nd}
\end{equation}
where $\hat{a}^\dagger_p$ ($\hat{a}_p$) creates (annihilates) an electron in MO $p$, and $h_{pq},\ h_{pqrs}$ are the one- and two-electron integrals, i.e.\ the expectation values of the one- and two-body operators in the first-quantized Hamiltonian \cref{ham1st} w.r.t.\ any possible combination of MOs.
A basis transformation can be done by reevaluating the electronic integrals using the new basis functions \cite{helgaker_molecular_2014,szabo_modern_1996}.

\subsubsection{Restricted, Unrestricted and Generalized Hartree--Fock}

In the original formulation of the Hartree--Fock equations the only restriction on HF wavefunctions was orthonormality \cite{fock_naherungsmethode_1930}. We can classify possible solutions based on the spin symmetries they preserve, a work first done by Fukutome \cite{fukutome_unrestricted_1981}.
The motivation behind this approach is that the Hamiltonian in \cref{eq:ham2nd} does not depend on the spin coordinate and thus commutes with the total spin operator squared $\hat S^2$, as well as the spin operator along the quantization axis $\hat S_z$. As a result, any eigenfunction of the Hamiltonian, i.e.\ an \textit{exact} solution to the Schrödinger equation, should also be an eigenfunction of those operators.
In contrast, any \textit{approximate} solution does not necessarily preserve the spin symmetry. If we enforce the conservation of a particular symmetry, the energy of the approximate solution increases. This ``symmetry dilemma" was first pointed out by Löwdin \cite{lefebvre_aspects_1969}.
Using the symmetry classification, we define the different HF solutions as follows:
The restricted HF (RHF) wavefunction is chosen to be an eigenfunction of both spin operators $\hat S^2$ and $\hat S_z$ (also called restricted open-shell HF (ROHF) for non-zero $\hat S_z$ eigenvalue), while in
unrestricted Hartree--Fock (UHF), the symmetry with $\hat S^2$ is broken. As a consequence, in the latter case spin contamination and spatial symmetry breaking could appear, even though the number of spin-up (or alpha spin) and spin-down (or beta spin) electrons are still fixed ($\hat S_z$ is conserved). It is also interesting to point out that, by projecting the symmetry-broken UHF state into different spin sectors, information about higher-level excitations can be obtained \cite{jimenez-hoyos_projected_2012}.
RHF often struggles with size-consistency problems \cite{jimenez-hoyos_generalized_2011}. UHF is often able to correct this issue and correctly dissociates a molecule to its fragments, but only if the open-shell electrons on a given fragment all have the same spin \cite{jimenez-hoyos_generalized_2011}. 
Finally, when relaxing all symmetry constraints, the (original) generalized Hartree--Fock (GHF) solution is obtained. 
This method allows for spin mixing in the molecular orbitals, i.e.\ MOs in GHF are no longer associated to a specific spin and $S_z$ is no longer conserved.

As a consequence, the GHF method can break spin symmetry to better represent systems with strong correlation, even in the dissociation limit. This method is also very useful for representing more difficult systems, such as molecules with diradical character or systems where magnetic interactions lead to conflicting spin alignments \cite{jimenez-hoyos_generalized_2011}.
Nevertheless, the application of GHF has remained rather limited, probably due to the fact that the resulting wave functions fail to include symmetries of the exact solution, and once good quantum numbers are lost, recovering them becomes challenging. In principle, these qualitative shortcomings may be addressed through post-HF correlation techniques, such as coupled cluster theory, although these approaches present their own difficulties \cite{jimenez-hoyos_generalized_2011}.

\subsubsection{Post-HF methods}\label{sec:Theo_2ndQ}

In the context of quantum chemistry, the Schrödinger equation can be solved exactly for a few systems only. In general, the exact solution can be obtained by finding the correct linear combination of all the determinants in the $N$-electron Fock space. 
As the molecular orbitals used to build the configuration states are limited by the selected atomic basis set, the exact solution is also restricted to the given orbital basis. 

There are two ways to find the exact coefficients of such a linear expansion. 
First, one can consider all possible combinations of Fock states $\{\vert\Psi_i\rangle\}$ to construct the matrices
\begin{align}
    \label{eq:hmat}
    H_{ij} &= \langle \Psi_i \vert \hat H \vert \Psi_j \rangle\\
    \label{eq:ovlpmat}
    S_{ij} &= \langle \Psi_i \vert \Psi_j \rangle
\end{align}
and find the generalized eigenvectors $\{\vec{v}\}$ and eigenvalues $\{E(\vec{v})\}$ such that
\begin{equation}\label{eq:gevp}
    H_{ij} \, \vec{v}_j = E(\vec v) S_{ij} \, \vec{v}_j.
\end{equation}
This procedure is called exact diagonalization. The diagonalization of the dense matrices yields all eigenstates and corresponding energies of the Hamiltonian. In practice, it is however only feasible to do a sparse diagonalization of the eigenvalues of interest.
Note that for orthonormal Fock states the overlap matrix becomes the identity, $S_{ij} = \delta_{ij}$.

Second, it is possible to find the coefficients variationally. In order to do so, an $N$-electron reference state, usually the HF state, is expanded by applying all possible excitation operators, thus creating all possible Fock states.
The coefficients of the expansion can be solved for by variational minimization of the energy, providing the exact ground state wave function, also known as full configuration interaction (FCI) solution. 

Unfortunately, the FCI method becomes rapidly unpractical because of the combinatorial increase in the number of Slater determinants related to the total number of spin orbitals in the given basis set and the number of electrons.

To generate classically feasible solutions, the expansion is usually truncated. The most extended method restricts the CI calculation to a selected set of spin orbitals, called the active space. This method is known as the complete active space configuration interaction (CASCI). Another possibility is to limit the order of the excitation operators used to build the configuration states. Generally, single and double excitations are the most important, making CI single and doubles (CISD) one of the most widely used methods of this kind.

\subsection{Active Space Selection}\label{sec:theory:active_space}
For our calculations in the active space (see \cref{sec:results:active_space}), we use a simple active space selection scheme minimizing jumps along the reaction coordinate (see also \cref{fig:casci_error} in the Appendix).
We start from the HF determinant and progressively add 2 additional spin orbitals based on their MO energy.
We consider two MOs as degenerate whenever their energy difference is smaller than $1.6 \,\mathrm{mE_h}$ (chemical accuracy). In that case, we use elements of the 2-body integral $h_{ijba}$ (see \cref{eq:ham2nd}) to estimate transition probabilities $T_\mathcal{A}(a,b)$ from all orbital pairs $i,j$ in the previous (smaller) active space $\mathcal{A}$ to the individual pairs $a,b$ in the set of degenerate orbitals as 
\begin{equation}
     T_\mathcal{A}(a,b) = \sum_{i,j\in\mathcal{A}} \vert\vert h_{ijba}\vert\vert.
\end{equation}
We then reorder the degenerate MOs starting with the orbitals having the largest $T_\mathcal{A}(a,b)$. For RHF only diagonal elements are considered where $a=b$ and $i=j$.

There are many other active space selection criteria in the literature \cite{stein_automated_2016, stein_autocas_2019, sayfutyarova_automated_2017, bao_automatic_2018, claudino_automatic_2019, kolodzeiski_automated_2023}. However, with our simple scheme, we can use the HF orbitals and do not require additional calculations. These are two advantages very appealing in the context of quantum computing and emphasize the importance of the initial HF calculation.

\subsection{Virtual Quantum Subspace Expansion}\label{sec:theory:VQSE}

Quantum subspace expansion methods construct a basis expansion around a reference state to improve expectation value accuracy and approximate excited states. In this work, we employ fermionic excitation operators to increase the energy accuracy of a reference wave function. Mathematically, we generate a set of basis functions $\vert \Psi_i \rangle$ from a common reference state $\vert \Psi_\text{REF} \rangle$ by applying operators ${\hat O}_i$,
\begin{equation}\label{eq:qse_basis}
    \vert \Psi_i \rangle = {\hat O}_i \vert \Psi_\text{REF} \rangle \,,
\end{equation}
and subsequently perform exact diagonalization within this subspace (\crefrange{eq:hmat}{eq:gevp}). Classical configuration interaction methods can be viewed as subspace expansions using a single-determinant Hartree-Fock state as the reference. In that context, the operator set $\mathcal O$ corresponds to the complete set of fermionic excitation operators for FCI, the complete set within an active space for CASCI, or the set of single and double excitations for CISD.

The key distinction of QSEs is the use of a multi-configurational reference state rather than a single-determinant Hartree-Fock state. Specifically, VQSE applies one-body and two-body fermionic operators that promote electrons from a multi-configurational active space solution to active or virtual orbitals. This approach is equivalent to the classical multi-reference CISD method.

We investigate three operator pools, the first two of which were introduced by Takeshita et al.~\cite{takeshita_increasing_2020}. All pools include operators defined as
\begin{equation}
    \mathcal{O} = \{\hat a_i^\dagger \hat a_p,\, \hat a_\mu^\dagger \hat a_q \hat a_\nu^\dagger \hat a_r\}\,.
\end{equation}
Virtual orbitals are unoccupied in the reference state by definition. Therefore, we restrict annihilation operators to the active space, denoted by $\mathcal{A}$, such that $p,q,r \in \mathcal{A}$. The pools are defined by restrictions on the creation indices:
\begin{itemize}
    \item \textbf{VQSE(V):} Single and double excitations are restricted to virtual orbitals ($i,\mu,\nu\in \mathcal{V}$).
    \item \textbf{VQSE(S+V):} Double excitations are restricted to virtual orbitals, while single excitations are allowed within the active space or to virtual orbitals ($i \in\mathcal{A} \cup \mathcal{V}$, $\mu,\nu\in \mathcal{V}$). This pool explicitly includes number operators $\hat{n}_p = \hat{a}_p^\dagger \hat{a}_p$ when $i=p$.
    \item \textbf{VQSE(all):} This pool contains all possible single and double excitations to active and virtual orbitals ($i,\mu,\nu \in\mathcal{A} \cup \mathcal{V}$). This includes products of number operators $\hat{n}_p \hat{n}_q$ when indices coincide.
\end{itemize}

\begin{figure}[!htb]
\centering
\includegraphics[width=\linewidth]{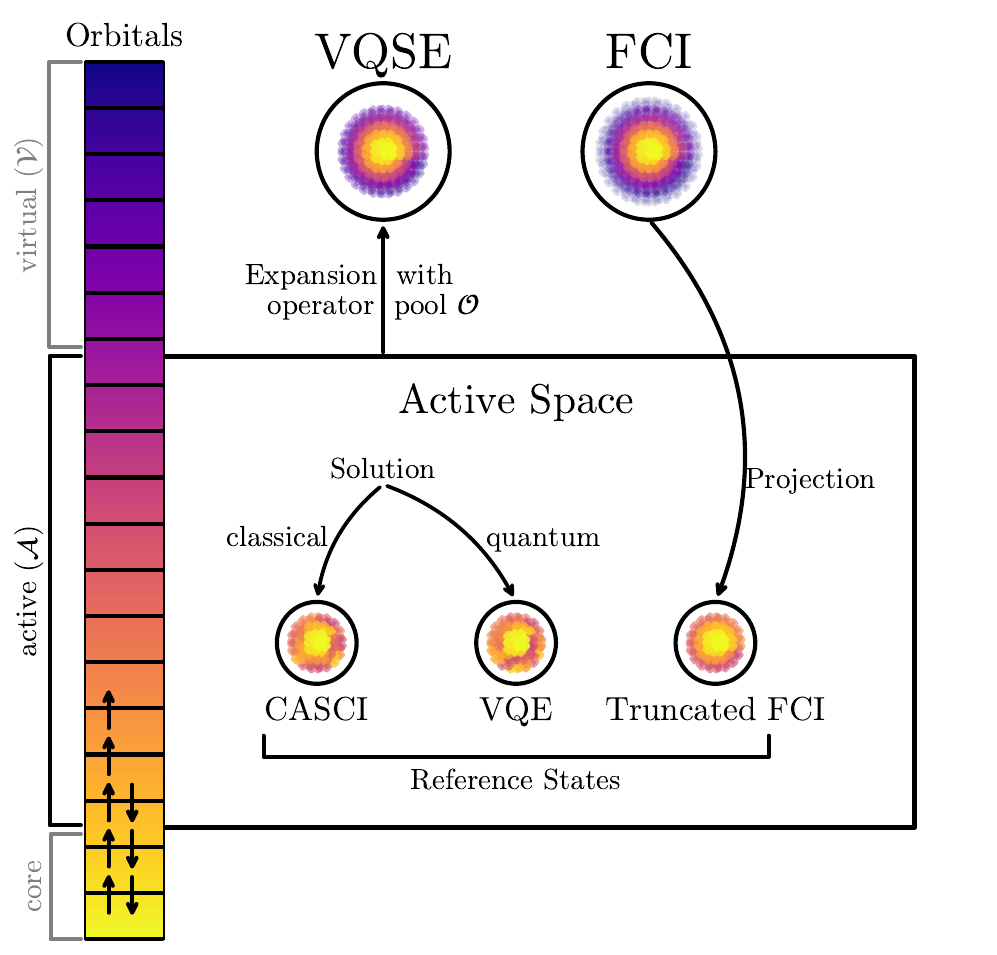}
\caption{Representation of VQSE reference states used in this work. After selecting active orbitals, we either approximately solve the CAS problem using CI or VQE or project the FCI solution to obtain Truncated FCI. VQSE expands any of these references inside the active space to recover FCI determinants.}
\label{fig:vqse_states}
\end{figure}

In principle, any function constructed within the active space serves as the reference state $\vert \Psi_\text{REF} \rangle$. In this work, we focus on three types of wave functions (see Fig.~\ref{fig:vqse_states}). First, the CASCI wave function, which is the exact solution to the Hamiltonian restricted to active space determinants. Second, the VQE solution using the UCCSD ansatz, which provides a variational approximation to the CASCI solution via energy minimization. Third, the truncated FCI wave function, obtained by projecting the exact FCI solution onto the Fock states of the active space and renormalizing the result. For all three reference types, the underlying molecular orbitals defining the active and virtual spaces are obtained from Hartree-Fock calculations (RHF, UHF, or GHF).

Generally, it is not necessary to prepare the basis functions $\vert \Psi_i \rangle$ explicitly. Instead we can insert the definition of \cref{eq:qse_basis} into \crefrange{eq:hmat}{eq:ovlpmat},
\begin{align}
    H_{ij} = \langle \Psi_i \vert \hat H \vert \Psi_j \rangle &= \langle \Psi_\text{ref} \vert \mathcal{\hat O}_i^\dagger \hat H \mathcal{\hat O}_j \vert \Psi_\text{ref} \rangle, \\
    S_{ij} = \langle \Psi_i \vert \Psi_j \rangle &= \langle \Psi_\text{ref} \vert \mathcal{\hat O}_i^\dagger \mathcal{\hat O}_j \vert \Psi_\text{ref} \rangle \,,
\end{align}
and then calculate the expectation values of the expansion operators $\mathcal{\hat O}_i^\dagger \mathcal{\hat O}_j$ and the transformed Hamiltonian $\mathcal{\hat O}_i^\dagger \hat H \mathcal{\hat O}_j$ with respect to the unchanged reference state. Therefore, the expansion can be performed on a quantum computer using only additional measurements.

An important implication of excluding double excitations to active orbitals is that after this transformation, by applying Wick's theorem, only terms with at most four creation and four annihilation operators contribute \cite{takeshita_increasing_2020}. It is therefore possible to construct the $H$ and $S$ matrices using up to the 4-RDM which scales as $N_A^8$, where $N_A$ is the number of active-space orbitals \cite{takeshita_increasing_2020}.

\subsubsection{GEVP Regularization}\label{sec:results:gevpr}

The generalized eigenvalue problem in VQSE requires regularization to filter out numerically unstable modes, i.e.\ to remove linearly dependent eigenvectors of the overlap matrix corresponding to zero eigenvalues. In the noiseless case, these eigenvalues are strictly null, allowing very small thresholds (e.g., $\epsilon=10^{-12}$) in numerical simulations, below which states are discarded. However, on noisy hardware, the regularization parameter $\epsilon$ must be chosen carefully. Following the work by Epperly et al. \cite{epperly_theory_2022}, we recommend selecting $\epsilon$ to be larger than, but close to, the expected hardware noise level. In our experiments on ibm\_aachen, we used $\epsilon = 10^{-2}$ based on calibration measurements of the readout error magnitude.
 
\section{Results}
\label{sec:results}

In the following subsections, we show our results for the dissociation of the hydrogen molecule. For reference, we first performed classical calculations, which allow us to evaluate the performance of VQSE on simulated noiseless and real noisy quantum hardware presented thereafter.

\subsection{Conventional Quantum Chemistry Calculations}
\label{sec:results:classical}

\subsubsection{Hartree-Fock and FCI}\label{sec:results:hf_fci}

We have calculated HF results with no unpaired electrons (singlet, $S_z=0$) and two unpaired electrons (triplet, $S_z=1$) for RHF, as well as GHF and singlet UHF. We have calculated the FCI results using the MOs of the different HF solutions and performed active space calculations with CASCI. This was done for a wide range of basis sets. 
For additional information we refer to \cref{appendix:classical} in the Appendix.

\begin{figure}[!ht]
\centering
\includegraphics[width=\linewidth]{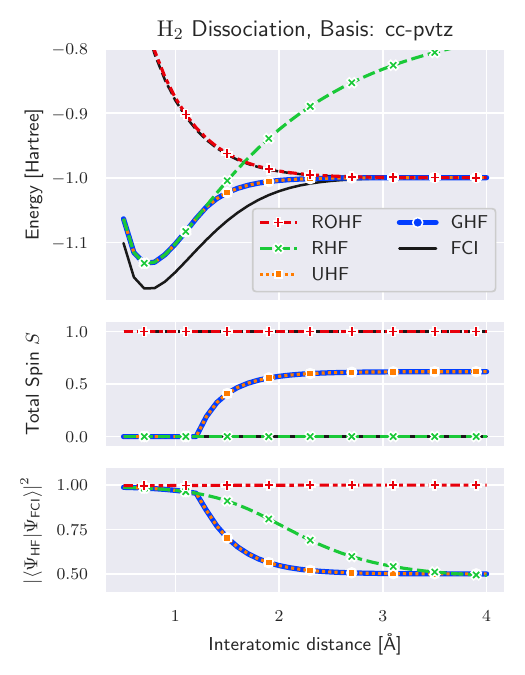}
\caption{
    $\mathrm H_2$ HF and FCI dissociation energy (top) in cc-pvtz basis, corresponding spin values (middle) and the overlap of HF with FCI (bottom) for singlet RHF (green), triplet ROHF (red), GHF (blue) and singlet UHF (orange). The FCI results are shown as black lines.
}
\label{fig:hf_fci}
\end{figure}

In Fig.~\ref{fig:hf_fci}, we show the $\mathrm{H}_2$ dissociation energies alongside total spin $S$ and overlap with FCI using the cc-pvtz basis. The FCI ground state is a singlet at equilibrium ($d_{\mathrm{H-H}} = 0.7$ \AA), becoming degenerate with the first excited, triplet state at large separation. ROHF closely matches the triplet FCI across all distances, while RHF fails to correctly model the FCI singlet. Its energy rises above the triplet state for distances larger than $\sim$\SI{1.7}{\angstrom}, an effect known as the singlet-triplet instability \cite{cizek_stability_1967}. 
To correctly describe dissociation, spin symmetry must be relaxed, yielding UHF and GHF. Near equilibrium, they match RHF; in the dissociation limit, they align with triplet ROHF. Between $\sim1.4$ \AA\ and $\sim2.8$ \AA, they smoothly connect equilibrium and dissociation, providing a qualitatively sound approximation of the singlet FCI \textit{energies}.
However, UHF and GHF exhibit spin contamination for intermediate and large distances, deviating from pure $S=0$ (singlet) or $S=1$ (triplet). The correct spin values are recovered with FCI, but we will later see that it requires a large amount of Slater determinants to do so (see \cref{fig:casci_nnz} in \cref{sec:results:active_space}). Notably, UHF and GHF give identical results here, unlike in other systems (e.g. $\mathrm{O}_2$\cite{jimenez-hoyos_generalized_2011}). As spin contamination increases, the UHF/GHF overlap with FCI drops rapidly, while the RHF overlap declines more slowly. All methods converge to a squared overlap of $\sim0.5$ at the dissociation limit.

\subsubsection{Active Space Analysis}\label{sec:results:active_space}

\begin{figure}[!ht]
    \centering
    \includegraphics[width=1\linewidth]{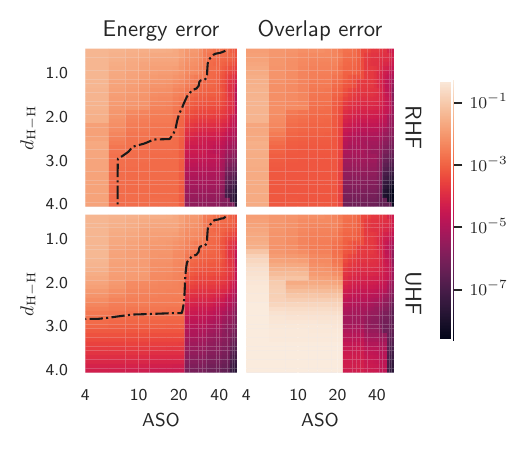}
    \caption{
        Energy error (left column) and overlap error (right column) of CASCI calculations with respect to FCI as a function of active spin orbitals (ASO) and bond distance for RHF (top row) and UHF (bottom row) in the cc-pvtz basis. The axes are arranged such that the dissociation limit and the FCI limit meet at the center, so that low errors converge toward the middle of the plot. Chemical accuracy (\SI{1.6e-3}{\hartree}) is indicated by a black contour line in the energy panels. GHF results are identical to UHF and omitted for clarity.
    }
    \label{fig:casci_heatmaps}
\end{figure}

The qualitative behavior observed in the HF solutions carries directly into active space calculations. When systematically increasing the number of active orbitals, \cref{fig:casci_heatmaps} shows the energy and overlap errors with respect to FCI as a function of active space size and bond distance for the RHF and UHF references in the cc-pvtz basis.
The two methods exhibit fundamentally different error profiles across the reaction coordinate: RHF maintains high overlap with the exact wavefunction even for small active spaces and large distances but requires more orbitals to reach chemical accuracy in the dissociation limit.
UHF models the correct dissociation energies with fewer orbitals but suffers from reduced overlap at intermediate and large distances due to spin contamination. GHF produces results identical to UHF for H$_2$ and is omitted for clarity.
The black contour line marks the active space sizes at which the chemical accuracy threshold is reached.
For the smallest distances up to 80\% of the total number of orbitals is needed and the number decreases for larger bond distances for all methods. This behaviour can be observed for all basis sets as shown in \cref{fig:supp-active_space} in the Appendix.
It is interesting to note that the usual practice of selecting the same amount of orbitals for active spaces along the reaction coordinate could lead to significant error in the energy differences as a consequence of different accuracies in the energy at different interatomic separations.

\begin{figure}[!ht]
    \centering
    \includegraphics[width=1\linewidth]{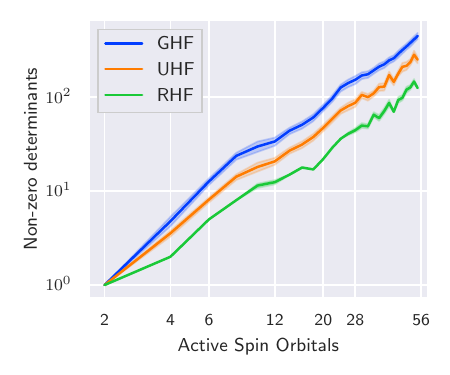}
    \caption{
        Number of non-zero determinants in the CASCI solution for the different HF methods as a function of active spin orbitals. RHF consistently requires fewer determinants than UHF or GHF to represent the CASCI wavefunction for a given active space size. The shaded area indicates the standard deviation introduced by averaging over bond distances.
    }
    \label{fig:casci_nnz}
\end{figure}

The reduced overlap of the UHF and GHF method means that more non-zero determinants are needed to describe the CASCI solution. This is shown in \cref{fig:casci_nnz}.

\subsection{Noiseless Quantum Simulations}
\label{sec:results:noiseless}

\subsubsection{VQE Parameter Analysis}\label{sec:results:vqe_params}

The number of non-zero variational parameters in the VQE calculation with the UCCSD ansatz corresponds to the number of non-zero excitation coefficients in the corresponding CASCI expansion (cf.~\cref{fig:casci_nnz}). We observe that UCCSD RHF requires fewer parameters than UHF or GHF near equilibrium, reflecting the simpler FCI expansion of the symmetry-adapted reference. However, for small active spaces in the dissociation limit the RHF energy error is larger, as discussed in \cref{sec:results:hf_fci}. For computational details of our VQE calculation, we refer to \cref{appendix:vqe} in the Appendix.

\subsubsection{VQSE Energy Convergence}\label{sec:results:vqse_energy}

\begin{figure*}[!ht]
\centering
\includegraphics[width=\textwidth]{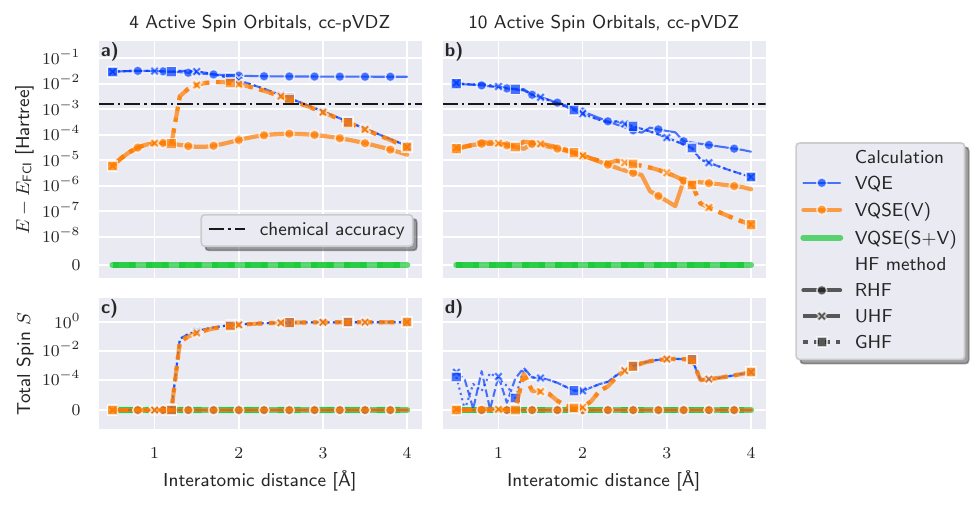}
\caption{a) and b): Increase of energy accuracy (difference to FCI). c) and d): Decrease in spin contamination (difference to 0) from VQE over VQSE(V) to VQSE(S+V) for the different HF methods. The left column (a) and c)) contains results for the smallest active space with 4 spin orbitals, the right column (b) and d)) contains data from the largest active space with 10 spin orbitals.
The chemical accuracy is indicated as a dashed-dotted line. Note that broken spin symmetry references (UHF/GHF) show improved VQE energies but may exhibit reduced VQSE(V) accuracy due to smaller FCI overlap.}
\label{fig:vqse_energy_vqe}
\end{figure*}

We perform a multivariate analysis of VQSE calculations based on
\begin{itemize}
    \item the size of the active space,
    \item the MOs (HF method) used,
    \item the operator pool (see \cref{sec:theory:VQSE}),
    \item and the reference state (see \cref{sec:theory:VQSE}).
\end{itemize}

We start by analyzing the VQSE energies from the VQE reference in the cc-pVDZ basis in \cref{fig:vqse_energy_vqe}. We compare the energies of the initial VQE calculation and the energies obtained with the operator pools VQSE(V) and VQSE(S+V) to the FCI energy. The VQSE(all) pool is equivalent to MR-CISD, which is equivalent to FCI for systems with two electrons. Therefore, we do not show results for this pool.
We show calculations for two active space sizes, at 4 (minimal) and 10 spin orbitals (half of the total number).
For further details we refer to \cref{appendix:vqse} in the Appendix.

As expected, the VQE energies are closer to FCI for the larger active space. Using UHF or GHF orbitals, we already achieve chemical accuracy for the smallest active space in the dissociation limit (distance $\geq 3\,$\AA).
The larger active space consistently reaches chemical accuracy already for distances greater than 1.7\,\AA\, for all HF methods. The broken spin symmetry methods, UHF and GHF, again always give the same result.

Turning to the VQSE results, we see a striking difference between the two operator pools. On the one hand, the set of expansion operators which includes operators inside the active space (VQSE(S+V)) is always able to exactly reproduce the FCI energy up to computational error.
The VQSE(V) operator pool, on the other hand, consistently reduces the energy error of the restricted references by 2 to 4 orders of magnitude, but never reaches FCI accuracy.
Additionally, for broken spin symmetry cases, this pool struggles to correct existing spin contamination (see panels c) and d)), leading to negligible reduction in energies around and after the spin crossing in the small active space. 
For the large active space, spin contamination around or below \SI{1e-4}{} is small enough to be corrected by the VQSE(V) pool, larger contamination remains unchanged.
However, the reduction in the energy error is not negatively affected and the results are comparable with the restricted case. At the dissociation limit of the large active space, the broken spin symmetry solution is even able to outperform the spin restricted case.

We can get a better understanding of these results by analyzing the overlap between the different references and the FCI state, and the spin.

\subsubsection{Overlap and Spin Analysis}\label{sec:results:overlap_spin}

\begin{figure}[!ht]
\centering
\includegraphics[width=\linewidth]{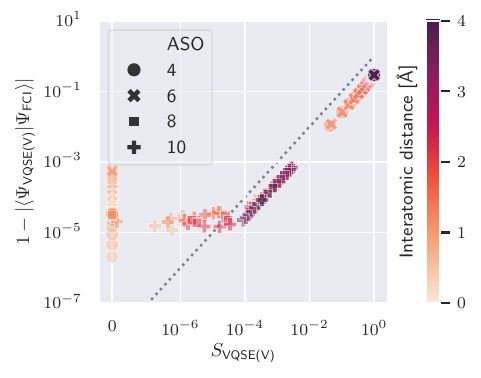}
\caption{Correlation between the overlap error and the spin contamination of VQSE(V) calculations at different distances with the cc-pvdz basis set.
The free spin VQE wavefunctions were used as reference states, and both UHF and GHF give very similar results with a standard deviation of \SI{1e-7}{}. The gray dotted line indicates perfect agreement.
}
\label{fig:spin_vs_overlap}
\end{figure}

In this section, we focus on the VQSE(V) pool, since VQSE(S+V) and VQSE(all) already provide perfect agreement with FCI.
Our results identify a fundamental constraint on the ability to modify active space determinants: the squared overlap between the reference state and the VQSE(V) wavefunction projected onto the active space and renormalized equals unity for all cases ($|\langle \Psi_\text{REF} \vert \Psi_{\text{VQSE(V)}\rightarrow\text{AS}} \rangle|^2 = 1$). This identity confirms that VQSE(V) preserves the relative coefficients of the active space determinants. Consequently, the fidelity of the VQSE(V) state is dictated linearly by the fidelity of the reference state. We observe this linear correlation across all references, active space sizes, and Hartree-Fock methods. Truncated FCI references enable exact FCI recovery because they inherently encode the correct relative active space coefficients by definition. In contrast, VQE and CASCI references possess inaccurate active space coefficients that VQSE(V) cannot relax, resulting in the finite energy errors observed in \cref{fig:vqse_energy_vqe}.

Unlike HF results, the overlap errors for VQE and CASCI references exhibit no clear dependence on bond distance. However, broken spin symmetry references (UHF and GHF) display increased spin contamination at larger bond distances, particularly for smaller active spaces. This contamination arises primarily from the mixing of configurations associated with the first excited (triplet) state. Because VQSE(V) preserves the relative coefficients of the active space determinants, it cannot eliminate these triplet contributions, resulting in persistent spin contamination.

Therefore, we observe a significant correlation between VQSE(V) overlap error and spin contamination when using UHF or GHF molecular orbitals, as shown in \cref{fig:spin_vs_overlap}. This correlation vanishes once spin contamination falls below \SI{1e-4}{}, a regime typically reached at smaller bond distances or with larger active spaces. These findings confirm that VQSE(V) accuracy is fundamentally limited by reference state fidelity. Consequently, mitigating spin contamination and improving energy accuracy requires operator pools capable of relaxing active space coefficients.

\subsection{Quantum Hardware Implementation}\label{sec:results_vqse_real}

\begin{figure*}[!htb]
    \centering
    \includegraphics[width=1\linewidth]{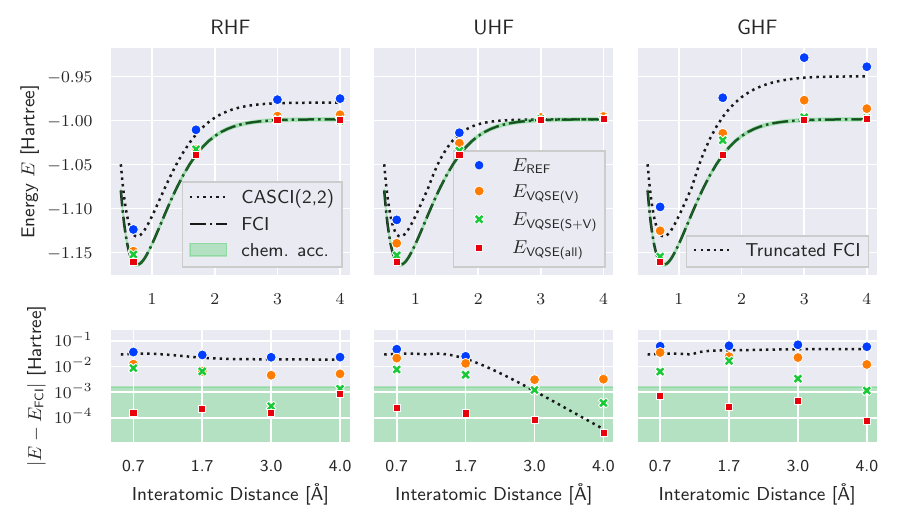}
    \caption{
    Energies of the reference states and VQSE evaluated on the ibm\_aachen quantum computer with the cc-pVDZ basis set. Results are shown for different operator pools VQSE(V), VQSE(S+V) and VQSE(all). The exact energy of the reference state is shown as a dotted line, the energy of the exact FCI solution is shown as the dashed-dotted line. The circuits used to construct the VQSE matrices were using CASCI (for RHF and UHF)/truncated FCI (for GHF) as a reference in the minimal active space with 4 spin orbitals, resulting in circuits with two/three qubits. The regularization procedure presented by Epperly et al.~\cite{epperly_theory_2022} is applied for the overlap matrix eigenvalues larger than \SI{1e-2}{}. }
    \label{fig:vqse_ibm_aachen}
\end{figure*}

To evaluate the performance on current quantum hardware, we ran the VQSE calculations on the IBM Aachen superconducting qubit device. We constructed circuits with the CASCI (RHF/UHF) and truncated FCI (GHF) wavefunction of the minimal active space (4 spin orbitals) in parity mapping using the Isometry ansatz \cite{iten_quantum_2016}, which yields circuits with two (RHF/UHF) and three (GHF) qubits. 
We prepared the reference state on the quantum computer and evaluated the energy ($E_\text{REF}$) as well as the VQSE energy ($E_\text{VQSE}$) using the same Pauli strings. Since the circuits are very shallow, we did not apply error mitigation. However, we filtered out purely imaginary expectation values and states with wrong particle number. Applying readout error mitigation, VQSE energies become more negative than FCI as the absolute value of Pauli string expectation values can exceed one. 

\cref{fig:vqse_ibm_aachen} compares the energies of the reference states, VQSE(V), VQSE(S+V), and VQSE(all) for RHF, UHF, and GHF orbitals under noisy conditions against FCI. In noiseless simulations, the VQSE(S+V) pool is sufficient to recover FCI accuracy regardless of the reference state. Under noisy conditions, however, VQSE(S+V) fails to mitigate spurious Slater determinants introduced to the reference by shot-noise and hardware errors, leading to significant energy deviations. Expanding to the full VQSE(all) pool improves accuracy by suppressing these intruder states, enabling chemical accuracy relative to FCI with the regularization threshold of \SI{1e-2}{}.
We find that only a select few operators account for the majority of the accuracy observed with VQSE(all) (cf.~\cref{sec:appendix:operator_selection}). This suggests that a targeted operator selection could avoid ill-conditioned GEVPs without relying on aggressive regularization, which stabilizes ground states but reduces spectral completeness by filtering out excited states \cite{kwao_generalized_2026}.
Noise also fundamentally alters the sensitivity to reference state quality compared to noiseless simulations. While noiseless VQSE(S+V) is robust across reference types, noisy simulations benefit significantly from reference states with lower variational energy. Consequently, UHF references consistently perform best under noisy conditions and truncated FCI no longer provides an advantage over CASCI. GHF requires an additional qubit and deeper circuits compared to RHF and UHF which renders it inefficient for noisy deployments. These findings highlight that successful VQSE calculations on noisy hardware require co-optimization of both operator pool expressivity and reference state efficiency.
\section{Discussion and Conclusion}
\label{sec:discussion_conclusion}

We examined how reference states and operator pools affect Virtual Quantum Subspace Expansion,
using H$_2$ dissociation as a benchmark. Our findings reveal that reference quality
operates through two mechanisms: overlap with the exact FCI wavefunction, and the expressivity of the VQSE operator pool.

Classical simulations establish several relevant trends. The two lowest-lying states of the hydrogen molecule have singlet and triplet spin multiplicity and become degenerate in the dissociation limit. The triplet is well described by the spin-restricted open-shell mean field solution. For the singlet, restricted solutions are inadequate in the transition and dissociated regions; only symmetry-broken UHF and GHF solutions capture the qualitatively correct energy curves. However, this improved energy accuracy from broken-symmetry references comes at the cost of reduced overlap with the FCI state. While they capture a larger amount of static correlation, the amount of dynamical correlation required to recover FCI also increases. Consequently, post-HF methods based on broken-symmetry orbitals require more determinants in the CI expansions or more parameters in VQE compared to the RHF case. For VQSE, it means more operators in the pool and, therefore, a numerically less well-conditioned GEVP. GHF did not outperform UHF and is more costly, as it uses determinants that are zero by construction for the other methods. In the absence of strong spin-manifold mixing or hardware noise, an RHF workflow is generally preferred for its superior overlap properties.

The choice of the VQSE operator pool significantly influences the accuracy of the results, and there is a strong co-dependency of operator pool and reference state.
The original study by Takeshita et al.~\cite{takeshita_increasing_2020} used a restricted operator pool (called VQSE(V) here) for numerical simulations, which produces small but finite energy errors (that we show to be further amplified by hardware noise, cf.~\cref{fig:vqse_ibm_aachen}). We find that by adding a small number of additional operators, namely, one-body operators within the active space [VQSE(S+V)], multi-reference CISD (MR-CISD) accuracy is recovered regardless of reference, which for H$_2$ is equivalent to the exact (FCI) solution.
In contrast, restricting excitation operators to act only from the active to the virtual space [VQSE(V)] renders results strongly dependent on the reference.  Because VQSE(V) does not modify coefficients of determinants within the active space, it cannot remedy reference-state deficiencies. Therefore, when using reduced operator pools, one should prioritize a reference state with large overlap to the FCI ground state rather than solely minimizing energy.

Under noisy conditions, operator pool requirements become more demanding. Using a regularization cutoff \cite{epperly_theory_2022} of \SI{1e-2}{} our results show that chemical accuracy relative to FCI can be achieved when using additional double excitation operators. The additional operators are crucial to filter out spurious Slater determinants introduced by noise but increase the theoretical measurement overhead from $N_A^{8}$ to $N_A^{12}$. Including many, linearly dependent operators reduces GEVP stability, but we found that only few operators actually contribute. These trade-offs motivate further investigation of operator pool design under realistic hardware constraints.

To monitor solution quality without FCI knowledge, we propose the total spin expectation value $\langle S^2 \rangle$ as a diagnostic. During dissociation, the ground state spin remains zero with no crossings. We find a strong correlation between overlap and spin errors, and therefore a correlation between energy and spin error,
enabling spin as a proxy for solution accuracy from noisy data. This diagnostic is informative only when the reference state permits spin contamination; spin-restricted references enforce symmetry and preclude this check. Thus, a trade-off exists between maximizing overlap and retaining benchmarking capability via spin.

Future work should focus on operator pools tailored for noise mitigation, as the present results demonstrate that hardware noise introduces qualitative requirements beyond those in noise-free settings. Combining such pools with reference states that balance static correlation capture and symmetry preservation, such as quantum-CASSCF \cite{yalouz_state-averaged_2021, tilly_reduced_2021, de_gracia_trivino_complete_2023, fitzpatrick_self-consistent_2024}, projected  wavefunctions~\cite{sugisaki_quantum_2019}, or approximate FCI solutions like Heat-Bath CI~\cite{holmes_heat-bath_2016} or DMRG~\cite{white_density_1992} truncated to the active space, may further improve VQSE performance on NISQ devices. Ultimately, successful deployment will require co-design of reference fidelity, pool expressivity, and hardware constraints. While H$_2$ is a simple benchmark, these findings motivate further investigation under realistic conditions.

\FloatBarrier

\begin{acknowledgments}
This project was made possible by the DLR Quantum Computing Initiative and the Federal Ministry for Economic Affairs and Climate Action; \url{qci.dlr.de/en/basiq}, as well as the Competence Center Quantum Computing Baden-Württemberg (KQCBW) through projects QuESt and QuESt+.
\end{acknowledgments}

\section*{Data Availability Statement}

The data that support the findings of this study are available from the corresponding authors upon reasonable request.

\clearpage
\newpage
\appendix
\section{Classical Calculations}\label{appendix:classical}

\begin{figure}[!ht]
\centering
\includegraphics[width=.95\linewidth]{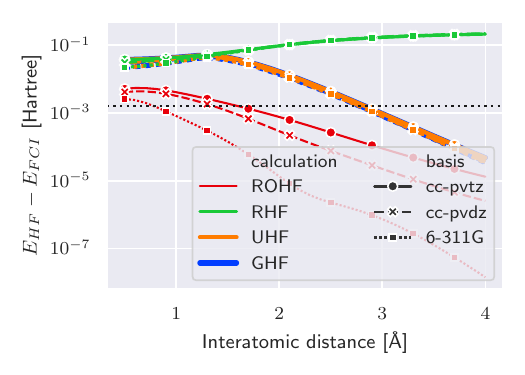}
\caption{
	Comparison of the amount of correlation energy gained by the singlet and triplet FCI calculations compared to the singlet (green), triplet (red) and broken spin symmetry (blue/orange) HF calculations. 
The quantum chemical accuracy is indicated as the dotted line.}
\label{fig:correlation_energy}
\end{figure}

\begin{figure}[!ht]
\centering
\includegraphics[width=\linewidth]{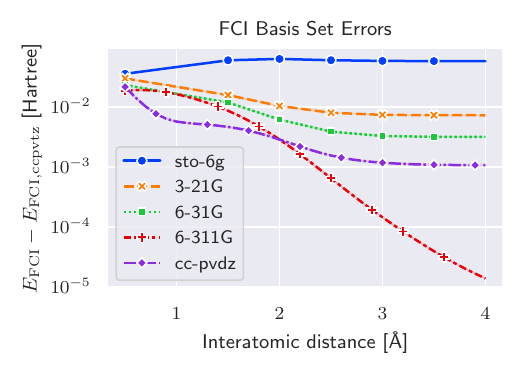}
\caption{
	$\mathrm H_2$ FCI groundstate error (difference to largest basis cc-pvtz) as a function of the interatomic distance
}
\label{fig:basis_set_error}
\end{figure}

We performed Hartree-Fock (HF) and post-HF calculations using the classical software \texttt{pySCF} (v 2.6) \cite{sun_pyscf_2018}. The energies were calculated at different fixed distances (between 0.5 and 4 \AA), until each self-consistent field (SCF) cycle converges and the energy change is below $10^{-11}$ hartrees. We used many different types of basis sets for the construction of the atomic orbitals, which are shown in \cref{fig:basis_set_error,fig:correlation_energy}. 

By choosing the type of HF calculation, that is, the spin symmetry of the electronic problem, we were able to obtain different shapes of the wave function between the restricted, unrestricted, and generalized cases (RHF, UHF, and GHF). However, this was not enough to obtain solutions of specific spins ($S$). According to the total number of electrons, the only two possibilities for H$_2$ are the singlet and triplet states, with spins $S = 0$ and $1$ and multiplicities $M_S = 1$ and $3$, respectively. One typical practice to get the desired spin is to select ROHF or UHF methods and set the initial number of unpaired electrons $n_s$ appropriately. It is important to note that the spin $S$ cannot be obtained directly counting $n_s$ but applying the spin operator $\hat{S^2}$ to the wave function. As mentioned before, only the RHF/ROHF method ensure the conservation of the spin (in this case $S = 0$/$S = 1$).  

As it is implemented in the \texttt{pySCF} code, we used the stability analysis on the HF solutions. This methodology provides a way to ensure a global minimum of the wave function in the space of HF solutions \cite{seeger_self-consistent_1977}. In UHF and GHF wave functions, finding the global minimum is crucial to break the spin symmetry.

From the HF solutions, we continued to calculate complete active space (CAS) and full configuration interaction (FCI) solutions. The FCI energies at different distances and with different basis sets were used as a reference because they correspond to the exact diagonalization of the complete Hamiltonian with the same number of particles and in the selected basis set. The spin of the CASCI and FCI wavefunctions was also calculated and compared to the one from HF solutions.

\begin{figure*}[!ht]
\centering
\includegraphics[width=\linewidth]{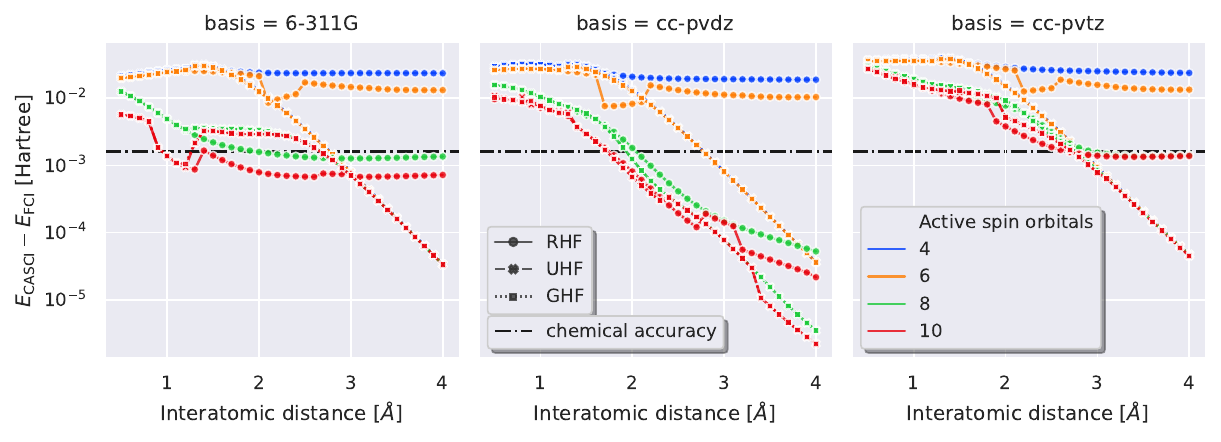}
\caption{Absolute energy error of $E_{\mathrm{AS}}$ relative to the respective FCI energies across varying basis sets, internuclear separations, and active space sizes. The chemical accuracy threshold is indicated by the black dash-dotted line.}
\label{fig:casci_error}
\end{figure*}

\begin{figure*}[!htb]
\centering
\includegraphics[width=\textwidth]{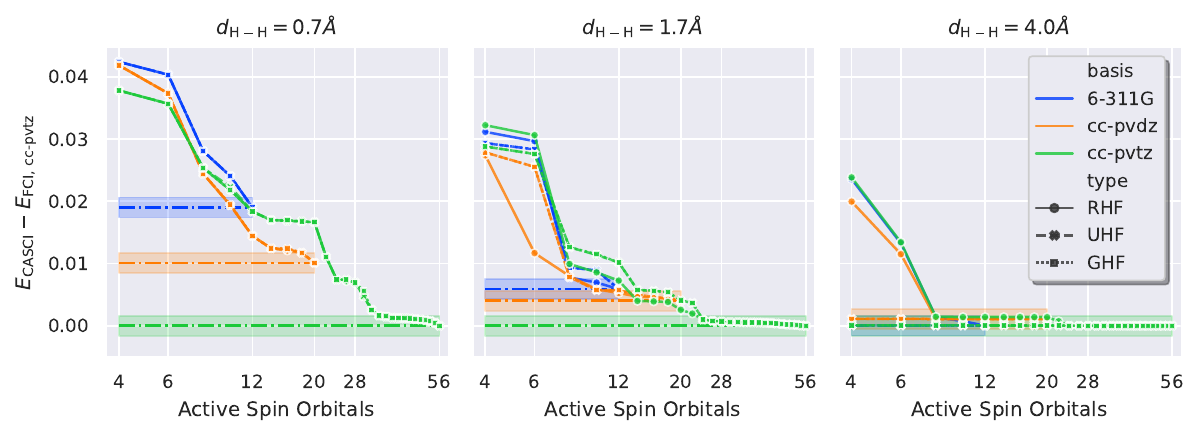}
\caption{Number of orbital necessary to achieve chemical accuracy to FCI using the different HF methods for the 6-311G, cc-pvdz and cc-pvtz basis sets. 
The FCI values are indicated with the colored dashed-dotted lines, and chemical accuracy to those values is indicated as the surrounding shaded area. The left panel shows the chemical bond 0.7 \AA,
the middle panel shows the spin crossing at 1.7 \AA, and the right panel shows the dissociation limit at 4.0 \AA. Energy values are shifted by the cc-pVTZ FCI value for each distance. 
}
\label{fig:supp-active_space}
\end{figure*}

Active spaces were created using canonical HF molecular orbitals. They were selected as described in \cref{sec:theory:active_space}. 
We have calculated CASCI solutions for the largest three basis sets, 6-311G, cc-pVDZ and cc-pVTZ. The results are shown in \cref{fig:supp-active_space,fig:casci_error}.

The energies of the CASCI calculations were used as a reference for the VQE calculations because they correspond to the exact diagonalization of the Hamiltonian projection in the AS.

\section{VQE Calculation}\label{appendix:vqe}

For the noiseless VQE calculations, the electronic Hamiltonian was constructed in second quantization using the electronic integrals calculated from the selected Hartree-Fock molecular orbitals.

We performed VQE calculations using version 4.2 of the \texttt{QuEST} simulator \cite{jones_quest_2019} in combination with the \texttt{SLSQP} optimizer from \texttt{SciPy} \cite{kraft1988software}.

We were using the UCCSD ansatz with second-order Trotter-Suzuki approximation \cite{Romero2019} and the Parity mapping with symmetry reduction \cite{bravyi_tapering_2017} as implemented in \texttt{qiskit} \cite{javadi-abhari_quantum_2024}.
The analytical gradient is used during optimization as described by Jones et al.\ \cite{jones_efficient_2020}.

\section{VQSE Calculation}\label{appendix:vqse}

The VQSE calculations were performed using an in-house algorithm written in rust with an interface to python for integration in the complete workflow.
Since the general structure of the VQSE matrices is the same for all studied cases, the fermionic operators are calculated only once for the largest active space, also leaving the coefficients of each hamiltonian element in a general form. Then for the smaller active spaces, Wicks Theorem is applied to rule out any operators evaluating to zero.
When it comes to mapping the fermionic operators, a similar procedure is followed. The parity mapping with 2-qubit reduction (RHF, UHF) and 1-qubit reduction (GHF) is calculated once for all unique fermionic operators in each active space.
The only step specific to the interatomic distance is then to calculate the expectation value of each unique pauli string with respect to the given reference state and inserting the hamiltonian coefficients.
Calculating the final VQSE matrices then boils down to simple lookup operations.
Using this optimized procedure, even the largest calculations with up to $\sim 12.5$ Million triagonal matrix elements could be performed in reasonable time on a conventional Laptop.

The expectation values of the unique pauli strings for each reference state are again evaluated with the \texttt{QuEST} simulator. The parity mapping and qubit reductions of the fermionic operators are performed using \texttt{qiskit} \cite{javadi-abhari_quantum_2024}. The encoding from Slater determinants to qubits of the CASCI and Truncated FCI reference wavefunctions is done manually.

\begin{figure*}[!ht]
\centering
\includegraphics[width=\textwidth]{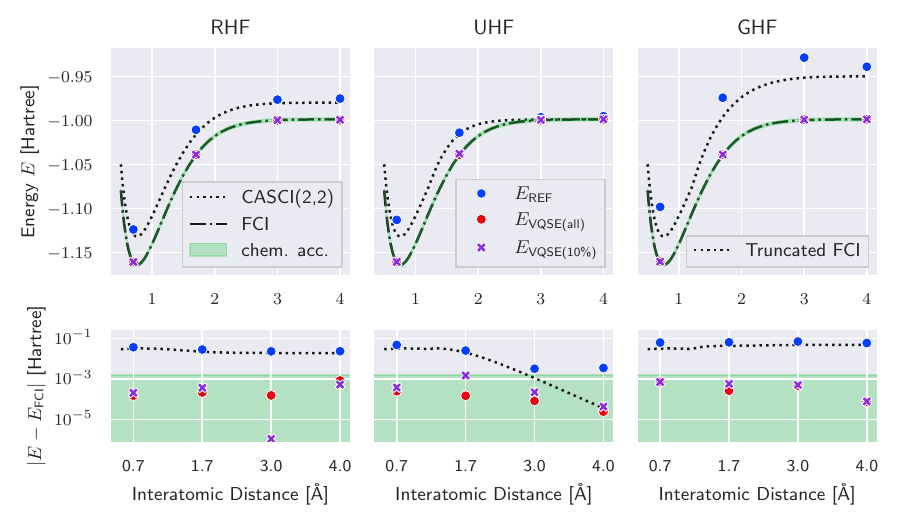}
\caption{Energies of the reference states and VQSE evaluated on the ibm\_aachen quantum computer with the cc-pVDZ basis set. Results are shown for the VQSE(all) operator pool and using 10\% of the most significant operators, which already achieves chemical accuracy for all cases. The exact energy of the reference state is shown as a dotted line, the energy of the exact FCI solution is shown as the dashed-dotted line. The circuits used to construct the VQSE matrices were using CASCI (for RHF and UHF)/truncated FCI (for GHF) as a reference in the minimal active space with 4 spin orbitals, resulting in circuits with two/three qubits. The regularization procedure presented by Epperly et al.~\cite{epperly_theory_2022} is applied for the overlap matrix eigenvalues larger than \SI{1e-2}{}. }
\label{fig:supp_noise_aware}
\end{figure*}

For the simulations on real hardware a circuit from the encoded wavefunction is created using the isometry ansatz\cite{iten_quantum_2016}. For measuring the different pauli expectation values, the necessary basis rotations are applied. Since we only use two or three qubits, we simply measure all unique, real pauli strings using qubit-wise commutation relations. Afterwards, the circuit is optimized and adapted to the target basis gates using the \texttt{qiskit} transpiler. \texttt{mapomatic} \cite{nation_suppressing_2023} is used to find the qubit layout with the lowest average error just before submitting the circuits.

After retrieving the measured bit strings, those corresponding to a wrong particle number (for GHF) are filtered out manually. 
We list the bit string counts as received by the ibm\_aachen computer in \cref{tab:bitstrings}.

\begin{table*}
    \centering
    \caption{Bit string counts for the hardware experiments as recieved by the ibm\_aachen quantum computer. Bit strings and Pauli strings are big-endian.}
\begin{tabular}{|c|c|c|c|c|c|c|}
\hline
MOs & $d_{\mathrm{H} - \mathrm{H}}$ & Pauli & \multicolumn{4}{c|}{Bitstring} \\\cline{4-7}
 & & & 00 & 01 & 10 & 11 \\\hline\hline
RHF & 0.7 & XX & 22799 & 27911 & 27422 & 21868 \\\hline
& & XZ & 383 & 50212 & 433 & 48972 \\\hline
& & YY & 22684 & 27933 & 27381 & 22002 \\\hline
& & ZX & 50020 & 49269 & 373 & 338 \\\hline
& & ZZ & 451 & 98816 & 424 & 309 \\\hline
& 1.7 & XX & 9581 & 40902 & 39085 & 10432 \\\hline
& & XZ & 5665 & 44868 & 6140 & 43327 \\\hline
& & YY & 10282 & 39867 & 40127 & 9724 \\\hline
& & ZX & 44294 & 44025 & 6078 & 5603 \\\hline
& & ZZ & 429 & 87972 & 11239 & 360 \\\hline
& 3.0 & XX & 1003 & 48750 & 49386 & 861 \\\hline
& & XZ & 19867 & 30643 & 20670 & 28820 \\\hline
& & YY & 989 & 48729 & 49330 & 952 \\\hline
& & ZX & 28790 & 29867 & 20819 & 20524 \\\hline
& & ZZ & 455 & 58165 & 41019 & 361 \\\hline
& 4.0 & XX & 582 & 49817 & 49017 & 584 \\\hline
& & XZ & 23713 & 26780 & 24817 & 24690 \\\hline
& & YY & 648 & 48880 & 49936 & 536 \\\hline
& & ZX & 25817 & 25645 & 25014 & 23524 \\\hline
& & ZZ & 432 & 51147 & 48073 & 348 \\\hline
UHF & 0.7 & XX & 22595 & 28147 & 27247 & 22011 \\\hline
& & XZ & 363 & 50173 & 387 & 49077 \\\hline
& & YY & 23391 & 27751 & 25832 & 23026 \\\hline
& & ZX & 49590 & 49658 & 404 & 348 \\\hline
& & ZZ & 422 & 98844 & 402 & 332 \\\hline
& 1.7 & XX & 42265 & 15481 & 18117 & 24137 \\\hline
& & XZ & 2897 & 54482 & 894 & 41727 \\\hline
& & YY & 32414 & 17712 & 16973 & 32901 \\\hline
& & ZX & 54292 & 41737 & 3238 & 733 \\\hline
& & ZZ & 753 & 95223 & 3196 & 828 \\\hline
& 3.0 & XX & 27321 & 23671 & 24744 & 24264 \\\hline
& & XZ & 423 & 49271 & 443 & 49863 \\\hline
& & YY & 25688 & 24261 & 24086 & 25965 \\\hline
& & ZX & 48812 & 50687 & 236 & 265 \\\hline
& & ZZ & 414 & 99109 & 140 & 337 \\\hline
& 4.0 & XX & 25187 & 24742 & 27039 & 23032 \\\hline
& & XZ & 456 & 50045 & 474 & 49025 \\\hline
& & YY & 25471 & 25125 & 24253 & 25151 \\\hline
& & ZX & 48349 & 51198 & 210 & 243 \\\hline
& & ZZ & 431 & 99129 & 108 & 332 \\\hline
\multicolumn{7}{c}{\vspace{17.7em}}
\end{tabular}\quad\begin{tabular}{|c|c|c|c|c|c|c|c|c|c|c|}\hline
MOs & $d_{\mathrm{H} - \mathrm{H}}$ & Pauli & \multicolumn{8}{c|}{Bitstring} \\\cline{4-11}
 &&& 000 & 001 & 010 & 011 & 100 & 101 & 110 & 111\\\hline\hline
GHF & 0.7 & XXX & 11808 & 13688 & 11829 & 13072 & 14559 & 10533 & 13937 & 10574 \\\hline
& & XXZ & 274 & 25670 & 288 & 24616 & 310 & 24837 & 339 & 23666 \\\hline
& & XYY & 12305 & 12306 & 12858 & 13079 & 12182 & 11847 & 12862 & 12561 \\\hline
& & XZX & 23242 & 26915 & 105 & 120 & 28222 & 21174 & 129 & 93 \\\hline
& & XZZ & 470 & 49818 & 26 & 189 & 602 & 48681 & 26 & 188 \\\hline
& & YXY & 11194 & 13862 & 11125 & 13407 & 14189 & 11439 & 13715 & 11069 \\\hline
& & YYX & 12745 & 11506 & 13192 & 12216 & 12648 & 11819 & 13230 & 12644 \\\hline
& & YYZ & 264 & 23759 & 277 & 25309 & 297 & 24270 & 325 & 25499 \\\hline
& & YZY & 22260 & 27137 & 96 & 101 & 27803 & 22384 & 117 & 102 \\\hline
& & ZXX & 26263 & 23908 & 25448 & 23629 & 200 & 185 & 186 & 181 \\\hline
& & ZXZ & 411 & 49855 & 437 & 48591 & 177 & 190 & 162 & 177 \\\hline
& & ZYY & 24180 & 24218 & 25530 & 25333 & 173 & 195 & 153 & 218 \\\hline
& & ZZX & 51237 & 47575 & 211 & 219 & 359 & 397 & 1 & 1 \\\hline
& & ZZZ & 783 & 98140 & 68 & 355 & 317 & 334 & 1 & 2 \\\hline
& 1.7 & XXX & 39293 & 1788 & 7901 & 1715 & 1759 & 37549 & 1627 & 8368 \\\hline
& & XXZ & 12737 & 28222 & 1443 & 8244 & 13384 & 25991 & 2012 & 7967 \\\hline
& & XYY & 14027 & 10065 & 10499 & 14648 & 10128 & 14857 & 14221 & 11555 \\\hline
& & XZX & 38172 & 3560 & 7105 & 175 & 3234 & 40660 & 279 & 6815 \\\hline
& & XZZ & 9657 & 31945 & 3485 & 3856 & 12839 & 31084 & 3531 & 3603 \\\hline
& & YXY & 37949 & 2453 & 7717 & 1943 & 1726 & 38757 & 1405 & 8050 \\\hline
& & YYX & 9983 & 15077 & 14249 & 10893 & 14219 & 10066 & 10938 & 14575 \\\hline
& & YYZ & 6893 & 18735 & 6336 & 18307 & 7893 & 16742 & 8505 & 16589 \\\hline
& & YZY & 38335 & 4633 & 6810 & 447 & 2697 & 39831 & 278 & 6969 \\\hline
& & ZXX & 27646 & 28789 & 7001 & 7625 & 12826 & 13535 & 1254 & 1324 \\\hline
& & ZXZ & 676 & 55638 & 561 & 14246 & 25903 & 426 & 2297 & 253 \\\hline
& & ZYY & 17356 & 18313 & 16167 & 19187 & 7470 & 6911 & 7530 & 7066 \\\hline
& & ZZX & 30697 & 32854 & 3968 & 3549 & 10600 & 11348 & 3488 & 3496 \\\hline
& & ZZZ & 933 & 62303 & 261 & 7501 & 21537 & 536 & 6729 & 200 \\\hline
& 3.0 & XXX & 21174 & 550 & 27639 & 671 & 775 & 20809 & 521 & 27861 \\\hline
& & XXZ & 9686 & 11849 & 12159 & 16306 & 10345 & 11350 & 13247 & 15058 \\\hline
& & XYY & 13575 & 11714 & 12409 & 11714 & 12134 & 13707 & 11623 & 13124 \\\hline
& & XZX & 47942 & 875 & 752 & 298 & 897 & 48185 & 380 & 671 \\\hline
& & XZZ & 21117 & 27506 & 403 & 690 & 23134 & 25972 & 530 & 648 \\\hline
& & YXY & 21028 & 970 & 27906 & 735 & 602 & 20676 & 489 & 27594 \\\hline
& & YYX & 11658 & 12959 & 11722 & 13713 & 12952 & 12291 & 13015 & 11690 \\\hline
& & YYZ & 11310 & 14862 & 11560 & 12744 & 10906 & 14328 & 11748 & 12542 \\\hline
& & YZY & 47856 & 1488 & 920 & 297 & 848 & 47557 & 289 & 745 \\\hline
& & ZXX & 11476 & 11499 & 15594 & 16056 & 9718 & 9837 & 13115 & 12705 \\\hline
& & ZXZ & 637 & 22792 & 439 & 31097 & 19023 & 607 & 24937 & 468 \\\hline
& & ZYY & 14256 & 14639 & 12478 & 13113 & 11422 & 10980 & 11667 & 11445 \\\hline
& & ZZX & 26574 & 26774 & 487 & 629 & 22361 & 22249 & 571 & 355 \\\hline
& & ZZZ & 779 & 52529 & 357 & 943 & 43739 & 732 & 581 & 340 \\\hline
& 4.0 & XXX & 26566 & 458 & 22181 & 326 & 809 & 26555 & 537 & 22568 \\\hline
& & XXZ & 11683 & 15278 & 9555 & 12644 & 13429 & 14308 & 11556 & 11547 \\\hline
& & XYY & 12392 & 13228 & 11651 & 12223 & 12764 & 12338 & 12682 & 12722 \\\hline
& & XZX & 50048 & 764 & 369 & 89 & 972 & 46649 & 147 & 962 \\\hline
& & XZZ & 21962 & 28955 & 217 & 233 & 22859 & 24663 & 518 & 593 \\\hline
& & YXY & 26156 & 1154 & 22034 & 476 & 702 & 26742 & 427 & 22309 \\\hline
& & YYX & 12977 & 12492 & 11737 & 12583 & 12887 & 12542 & 12565 & 12217 \\\hline
& & YYZ & 11338 & 14161 & 10317 & 14099 & 12183 & 13250 & 12187 & 12465 \\\hline
& & YZY & 47868 & 1284 & 488 & 322 & 914 & 48538 & 210 & 376 \\\hline
& & ZXX & 14583 & 13429 & 13104 & 12815 & 12335 & 11034 & 11640 & 11060 \\\hline
& & ZXZ & 420 & 27971 & 552 & 25454 & 22967 & 378 & 21639 & 619 \\\hline
& & ZYY & 13061 & 14314 & 12390 & 14550 & 12205 & 11195 & 11874 & 10411 \\\hline
& & ZZX & 27409 & 26001 & 326 & 526 & 23425 & 21597 & 244 & 472 \\\hline
& & ZZZ & 622 & 52590 & 246 & 601 & 44479 & 732 & 495 & 235 \\\hline
    \end{tabular}
    \label{tab:bitstrings}
\end{table*}

\subsection{Solving the GEVP}

To solve the GEVP, we calculate the eigenvalues and eigenvectors of the overlap matrix and arrange the eigenvectors with eigenvalues greater than the threshold as columns of a transformation matrix $P$. We then calculate the transformed matrix $\tilde{H} = \Lambda^{-1}P^T H P$, where $\Lambda$ is a diagonal matrix of the selected eigenvalues. We perform a sparse diagonalization of $\tilde H$ to obtain transformed eigenvectors $\tilde{\mathbf v}$ and energies. The eigenvectors w.r.t.~the full matrix can be obtained by calculating $\mathbf{v} = P \tilde{\mathbf v}$.

\subsection{Operator Truncation for Improved Scaling}\label{sec:appendix:operator_selection}

To assess the potential for improved computational scaling, we evaluated the algorithm's performance using a truncated subset of dominant operators. Specifically, we ranked the operators based on the magnitude of their corresponding coefficients in the ground state eigenvector $\mathbf{v}$ and retained only the top 10\%. The resulting energies, denoted as $E_\mathrm{VQSE(10\%)}$, are presented in \cref{fig:supp_noise_aware} alongside the results when using the full operator pool,$E_\mathrm{VQSE(all)}$. Notably, chemical accuracy is preserved across all test cases despite the 90\% reduction in operators, highlighting the potential for improved computational scaling.

\FloatBarrier
\clearpage

\bibliography{references}
\end{document}